\documentclass{kapproc}
\usepackage{t1enc} 
\usepackage{procps}
\usepackage[dvips]{graphicx}
\usepackage{amssymb} 
\usepackage{epsf}    
\upperandlowercase
\kluwerbib

\begin{document}

\articletitle[Modes of star formation along the Hubble Sequence and
beyond]{Modes of star formation along the\\ Hubble Sequence and
beyond}

\author{Richard de Grijs}
\affil{Department of Physics \& Astronomy, The University of
Sheffield, Hicks Building, Hounsfield Road, Sheffield S3 7RH, UK}

\begin{abstract}
I assess the similarities and differences between the star-formation
modes in quiescent spiral galaxies versus those in violent starburst
regions. As opposed to the quiescent star-formation mode in spiral
galaxies, current empirical evidence on the star-formation processes
in the extreme, high-pressure environments induced by galaxy
encounters strongly suggests that star {\it cluster} formation is an
important and perhaps even the dominant mode of star formation in such
starburst events. The sizes, luminosities, and mass estimates of the
young massive star clusters (YMCs) are entirely consistent with what
is expected for young Milky Way-type globular clusters (GCs). Recent
evidence lends support to the scenario that GCs, which were once
thought to be the oldest building blocks of galaxies, are still
forming today. One of the key unanswered questions in this field
relates to their possible survival chances for a Hubble time, and thus
to the potential evolutionary connection between YMCs and GCs.
\end{abstract}

\section{Quiescent versus violent modes of star formation}

Star formation plays a major role in the evolution of galaxies.
However, despite this, it is still rather poorly understood. In fact,
on galaxy-wide scales, we do not have much more than a few scaling
relations -- such as the Schmidt-Kennicutt law (Schmidt 1959,
Kennicutt 1998) -- at our disposal. The process of star formation
itself, while central to galaxy astrophysics, is one of the remaining
``dark corners of the evolutionary process'' (e.g., O'Connell 2005).

The Schmidt-Kennicutt law, in one of its modern guises relating the
(galactic) star formation density to the gas surface density (e.g.,
Kennicutt 1998), is remarkably successful in describing the
star-formation properties of galaxies ranging from quiescent spirals
(such as the Milky Way), via circumnuclear and localised star-forming
environments (e.g., M83, NGC 6946), to major, galaxy-wide bursts of
star formation (e.g., M82, the ``Antennae'' system) and -- in the
extreme -- the ultraluminous infrared galaxies (ULIRGs) thought to be
the remnants of major galaxy-galaxy interactions. The global
star-formation rates in these wide-ranging environments correlate very
closely with the local gas surface density over at least 5 orders of
magnitude.

While this suggests very strongly that star formation is indeed driven
by the local gas surface density, I point out that in order to
maintain the high star-formation rates seen in starburst environments,
the star-formation {\it efficiency} needs to be maintained at moderate
to high levels for a significant length of time. Star-formation
efficiencies, defined here as the total mass in stars with respect to
the total available gas mass in a given environment, range from $0.1 -
3$ per cent for quiescent spiral and irregular galaxies, as well as
for dwarf starbursts, to $\gtrsim 10-30$ per cent in the more extreme
environments of major interactions and the most violently star-forming
areas in ULIRGs (e.g., Gao \& Solomon 2004a,b).

\section{Star formation in interacting galaxies}

Perhaps not surprisingly, the highest star-formation efficiencies are
generally found in the context of gravitationally interacting
galaxies. [For the purpose of this discussion I ignore any possible
connection between active galactic nuclei and starbursts.] Many
authors have pointed out that such major star-forming episodes are
difficult, if not impossible, to sustain for any cosmologically
significant length of time, as the high star-formation efficiencies
will deplete the available gas reservoir on short time-scales
(generally, $\lesssim 10^7$ yr), quickly lead to massive supernova
explosions, and thus quench the ongoing starburst (e.g., Chevalier \&
Clegg 1985; Doane \& Mathews 1993; see also, e.g., Schweizer et
al. 1996). The limitations on the duration of most starburst events
imposed by the available gas content imply that most ongoing
starbursts cannot have sustained their high levels of active star
formation for significant lengths of time. These considerations
naturally suggest that galaxy interactions (or equivalent events) and
enhanced levels of active star formation are intimately linked;
conversely, starburst events are often (but not exclusively) linked to
ongoing or recent (generally within the last $\sim 10^{8-9}$ yr)
galaxy mergers or interactions.

\subsection{The importance of star {\it cluster} formation}

It is becoming increasingly clear that a major fraction of the star
formation in interacting galaxies likely takes place in the form of
dense, massive star clusters (e.g., de Grijs et al. 2001, 2003d, and
references therein). In de Grijs et al. (2003d) we analysed the {\sl
Hubble Space Telescope (HST)}/ACS Early Release Observations of the
``Mice'' and the ``Tadpole'' interacting systems, using a
pixel-by-pixel approach. In both systems we found more than 40 young
massive star-forming regions (possibly star clusters, although the
distances to these galaxies of $\sim 100$ Mpc precluded us from
concluding so more robustly) -- we noted that these most likely
represented the proverbial ``tip of the iceberg'' because of
observational selection effects. Overall, we found that more than 35
per cent of the active star formation in these galaxies occurred in
the dense star-forming regions; in the tidal tails and spiral arms, at
least 75 per cent of the blue ($B$-band) flux originated from these
areas.

This preponderance of dense star-forming ``clusters'' can be
understood easily in the context of the previous discussion. Elmegreen
\& Efremov (1997) showed, for instance, that in order to produce a
massive star ``cluster'' -- for a give ambient pressure -- a high
star-formation efficiency is required. Their figure 4 clearly shows
that the more massive progenitor molecular clouds, with masses in
excess of $\sim 10^7$ M$_\odot$, are more conducive to high
star-formation efficiencies than their lower-mass counterparts.  In
the presence of significant external pressure (as in the case of
interacting galaxies), and high star-formation efficiencies (as found,
once again, in interacting galaxies), the resulting star formation
through cloud collapse is therefore expected to occur predominantly in
the form of massive star-forming regions.

\subsection{Decoupled star and cluster formation}

This scenario is qualitatively supported by the detailed star and star
cluster formation histories that have been derived for a number of
nearby galaxies. In M82's fossil starburst region ``B'', for instance,
we found a clear and dominant peak in the star cluster formation
history around $10^9$ yr ago (de Grijs et al. 2001, 2003a). However,
star cluster formation declined rapidly (within a few $\times 10^8$
yr) after this burst episode, while quiescent star formation in the
galaxy's disk continued until at least 20 to 30 Myr ago (de Grijs et
al. 2001). This indicates a physical decoupling between the star and
star cluster formation modes in M82 B, implying special conditions as
prerequisite for star cluster formation (see also Ashman \& Zepf 1992;
Elmegreen \& Efremov 1997).

This situation is, in fact, similar to that in the Large Magellanic
Cloud (LMC), and also to that in the nearby starburst dwarf galaxy NGC
5253. In the LMC, the star cluster population consists of a dozen old,
globular cluster (GC)-like objects with ages of $\sim 13$ Gyr, and a
large complement of ``populous'' clusters, all younger than $\sim 3-4$
Gyr. With the exception of a single object, ESO 121-SC03, there are no
known populous clusters in the LMC with robustly confirmed ages in the
``age gap'' between about 3 and 13 Gyr (see, however, Rich et
al. 2001, and references therein). The LMC's field star population, on
the other hand, does not show any such marked absence of
star-formation activity in this period, in neither the central bar
region nor the outer disk field (see, e.g., Holtzman et al. 1999).

Tremonti et al. (2001) obtained high-resolution {\sl Hubble Space
Telescope}/ STIS spectroscopy of both a number of young massive star
clusters (YMCs) and the diffuse field population in the disk of NGC
5253. Their composite cluster spectrum shows strong O-star signature
(P Cygni profiles associated with massive stellar winds) that are not
seen in the composite field-star spectrum. This raises the question,
once again, whether star formation might operate differently in the
field than in the clusters. They concluded that a fair fraction of the
field star population might consist of dissolved star clusters. If
this is the correct interpretation, this would imply that most of the
YMCs in NGC 5253 dissolve on time-scales of $\sim 10$ Myr. It has
recently been pointed out that this type of ``infant mortality'' might
be plausible in other interacting galaxies such as the Antennae
(Whitmore 2004) and M51 (Bastian et al. 2005) as well.

\section{Fate of the young massive star clusters}

Despite the infant mortality process observed in some of the nearest
interacting systems, significant fractions of star clusters induced by
the interactions survive to become bound objects. This is supported,
e.g., by the existence of $\sim 1$ Gyr-old clusters in M82 B (de Grijs
et al. 2001, 2003a), thought to have been triggered by its last tidal
encounter with M81, and other galaxies containing star clusters older
than a few $\times 10^7$ yr.

In many cases, these clusters resemble young {\it globular} clusters
based on their current sizes, luminosities and photometric (and
sometimes spectroscopic) mass estimates. The question remains,
however, whether at least a fraction of these young objects will
survive to ages of $\gtrsim 10$ Gyr. GCs have long been thought to be
the oldest building blocks of galaxies, formed at or before the time
of galaxy formation. Yet, the discovery of YMCs resembling GC
progenitors begs the question whether we are witnessing the formation
of proto-GCs in the most violent star-forming regions. If we can prove
this conclusively, one way or the other, the answer to this crucial
question will have far-reaching consequences for our understanding of
a wide range of astrophysical issues, with bearing on the key issues
of the nature of star formation, galaxy assembly and evolution.

There are a number of approaches to address this question. Here, I
will focus on the use of the cluster luminosity (or mass) function
(CLF; i.e., the number of clusters in a population within a given
luminosity or mass range). The seminal work by Elson \& Fall (1985) on
the young LMC cluster system (with ages $\lesssim 2 \times 10^9$ yr)
seems to imply that the cluster luminosity function (CLF) of YMCs is
well described by a power law of the form $N_{\rm YSC}(L) {\rm d} L
\propto L^{\alpha} {\rm d} L$, where $N_{\rm YSC}(L) {\rm d} L$ is the
number of YSCs with luminosities between $L$ and $L + {\rm d} L$, with
$-2 \lesssim \alpha \lesssim -1.5$ (see de Grijs et al. 2003c for a
review). On the other hand, for old GC systems with ages $\gtrsim
10^{10}$ yr, the CLF shape is well established to be roughly Gaussian
(or log-normal), characterized by a peak (turn-over) magnitude at
$M_V^0 \simeq -7.4$ mag and a Gaussian FWHM of $\sim 3$ mag (e.g.,
Whitmore et al. 1995; Harris et al. 1998).

This type of observational evidence has led to the popular, but thus
far mostly speculative theoretical prediction that not only a
power-law, but {\it any} initial CLF (or cluster mass function; CMF)
will be rapidly transformed into a Gaussian (or log-normal)
distribution because of (i) stellar evolutionary fading of the
lowest-luminosity (and therefore lowest-mass) objects to below the
detection limit; and (ii) disruption of the low-mass clusters due both
to interactions with the gravitational field of the host galaxy, and
to internal two-body relaxation effects leading to enhanced cluster
evaporation (e.g., Elmegreen \& Efremov 1997; Gnedin \& Ostriker 1997;
Ostriker \& Gnedin 1997; Fall \& Zhang 2001).

\subsection{The M82 B cluster population: initial conditions revealed?}

In de Grijs et al. (2003a,b) we reported the discovery of an
approximately log-normal CLF (and CMF) for the roughly coeval star
clusters at the intermediate age of $\sim 1$ Gyr in M82 B. This
provided the first deep CLF (CMF) for a star cluster population at
intermediate age, which thus serves as an important benchmark for
theories of the evolution of star cluster systems. Recently, we
further investigated whether the most likely initial CMF in M82 B was
more similar to either a log-normal or a power-law distribution (de
Grijs et al. 2005), by taking into account the dominant evolutionary
processes (including stellar evolution, and internal and external
gravitational effects) affecting the mass distributions of star
cluster systems over time-scales of up to $\sim 1$ Gyr in the presence
of a realistic underlying gravitational potential.

From our detailed analysis of the expected evolution of CMFs, we
conclude that our observations of the M82 B CMF are inconsistent with
a scenario in which the 1 Gyr-old cluster population originated from
an initial power-law mass distribution. This applies to a range of
characteristic disruption time-scales. Our conclusion is supported by
arguments related to the initial density in M82 B, which would be
unphysically high if the present cluster population were the remains
of an initial power-law distribution.

In a static gravitational potential, which is a good approximation to
the M82 B situation despite its recent encounter with M81 (see de
Grijs et al. 2005), Vesperini (1998) shows conclusively that there
exists a particular CMF of which the initial mean mass, width and
radial dependence remain unaltered during the entire evolution over a
Hubble time. In fact, the mean mass and width of {\it any} initial
log-normal CMF tends to evolve towards the values for this equilibrium
CMF, which is very close (within the 1$\sigma$ uncertainties) to the
current parameterisation of the M82 B CMF. By extension, this may
imply that we are indeed witnessing the formation of proto-globular
clusters in M82 -- provided that a fraction of them survive for a
Hubble time.

\begin{acknowledgments}
I thank, in particular but in no particular order, Uta
Fritze--v. Alvensleben, Genevi\`eve Parmentier, Peter Anders, Henny
Lamers and Nate Bastian for their valuable comments on and
contributions to various aspects of the studies highlighted in this
article. This work has been partially supported by the International
Space Science Institute in Berne (Switzerland).
\end{acknowledgments}


\begin{chapthebibliography}{}
\bibitem[]{} Ashman K.M., Zepf S.E., 1992, ApJ, 384, 50
\bibitem[]{} Bastian N., Gieles M., Lamers H.J.G.L.M., Scheepmaker R.,
de Grijs R., 2005, A\&A, 431, 905
\bibitem[]{} Chevalier R.A., Clegg A.W., 1985, Nature, 317, 44
\bibitem[]{} de Grijs R., O'Connell R.W., Gallagher J.S. {\sc iii},
2001, AJ, 121, 768
\bibitem[]{} de Grijs R., Bastian N., Lamers H.J.G.L.M., 2003a, MNRAS,
340, 197
\bibitem[]{} de Grijs R., Bastian N., Lamers H.J.G.L.M., 2003b, ApJ,
583, L17
\bibitem[]{} de Grijs R., Anders P., Lynds R., Bastian N., Lamers
H.J.G.L.M., O'Neill E.J., Jr., 2003c, MNRAS, 343, 1285
\bibitem[]{} de Grijs R., Lee J.T., Mora Herrera M.C., Fritze--v.
Alvensleben U., Anders P., 2003d, New Astron., 8, 155
\bibitem[]{} de Grijs R., Parmentier G., Lamers H.J.G.L.M., 2005,
MNRAS, submitted
\bibitem[]{} Doane J.S., Mathews W.G., 1993, ApJ, 419, 573
\bibitem[]{} Elmegreen B.G., Efremov Y.N., 1997, ApJ, 480, 235
\bibitem[]{} Fall S.M., Zhang Q., 2001, ApJ, 561, 751
\bibitem[]{} Gao Y., Solomon P.M., 2004a, ApJ, 606, 271
\bibitem[]{} Gao Y., Solomon P.M., 2004b, ApJS, 152, 63
\bibitem[]{} Gnedin O.Y., Ostriker J.P., 1997, ApJ, 474, 223
\bibitem[]{} Harris W.E., Harris G.L.H., McLaughlin D.E., 1998, AJ, 115,
1801
\bibitem[]{} Holtzman J.A., et al., 1999, AJ, 118, 2262
\bibitem[]{} Kennicutt R.C., Jr., 1998, ARA\&A, 36, 189
\bibitem[]{} O'Connell R.W., 2005, in: Starbursts -- from 30 Doradus
to Lyman break galaxies, Astrophysics \& Space Science Library, de
Grijs R., Gonz\'alez Delgado R.M., eds., Springer: Dordrecht, ASSL,
329, 333
\bibitem[]{} Ostriker J.P., Gnedin O.Y., 1997, ApJ, 487, 667
\bibitem[]{} Rich R.M., Shara M.M., Zurek D., 2001, AJ, 122, 842 
\bibitem[]{} Schmidt M., 1959, ApJ, 129, 243
\bibitem[]{} Schweizer F., Miller B.W., Whitmore B.C., Fall, S.M., 1996,
AJ, 112, 1839
\bibitem[]{} Tremonti C.A., Calzetti D., Leitherer C., Heckman T.M.,
2001, ApJ, 555, 322
\bibitem[]{} Vesperini E., 1998, MNRAS, 299, 1019
\bibitem[]{} Whitmore B.C., Sparks W.B., Lucas R.A., Macchetto F.D.,
Biretta J.A., 1995, ApJ, 454, L73
\bibitem[]{} Whitmore B.C., 2004, in: The Formation and Evolution of
Massive Young Star Clusters, ASP Conf. Ser., vol. 322, Lamers
H.J.G.L.M., Smith L.J., Nota A., eds., ASP: San Francisco, p. 419


\end{chapthebibliography}
\end{document}